Title: Computational Mechanism for the Effect of Psychosis Community Treatment: A Conceptual Review from Neurobiology to Social Interaction

David Benrimoh, Andrew Sheldon[2], Ely Sibarium[2], Albert R. Powers[2]

[2] Department of Psychiatry, Yale University School of Medicine, New Haven, CT USA

ABSTRACT

The computational underpinnings of positive psychotic symptoms have recently received significant attention. Candidate mechanisms include some combination of maladaptive priors and reduced updating of these priors during perception. A potential benefit of models with such mechanisms is their ability to link multiple levels of explanation, from the neurobiological to the social, allowing us to provide an information processing-based account of how specific alterations in self-self and self-environment interactions result in the experience of positive symptoms. This is key to improving how we understand the experience of psychosis. Moreover, it points us towards more comprehensive avenues for therapeutic research by providing a putative mechanism that could allow for the generation of new treatments from first principles. In order to demonstrate this, our conceptual paper will discuss the application of the insights from previous computational models to an important and complex set of evidence-based clinical interventions with strong social elements, such as coordinated specialty care clinics (CSC) in early psychosis and assertive community treatment (ACT). These interventions may include but also go beyond psychopharmacology, providing, we argue, structure and predictability for patients experiencing psychosis. We develop the argument that this structure and predictability directly counteract the relatively low precision afforded to sensory information in psychosis, while also providing the patient more access to external cognitive resources in the form of providers and the structure of the programs themselves. We discuss how computational models explain the resulting reduction in symptoms, as well as the predictions these models make about potential responses of patients to modifications or to different variations of these interventions. We also link, via the framework of computational models, the patient's experiences and response to interventions to putative neurobiology.

INTRODUCTION:

While positive symptoms- such as hallucinations and delusions- have not been demonstrated to be the chief determinant of functional outcomes (Bowie and Harvery, 2006), their appropriate management, for many people on the psychotic spectrum, do appear to be of significant importance in terms of quality of life, vocational functioning, and safety (Wiersma et al., 2004; Shawyer et al., 2003; Mucci et al., 2021). Hallucinations and delusions may cause fear and distress, occupy a significant portion of a patient's attention and time that could be spent on other tasks, or lead to unsafe behavior such as avoiding necessary medical treatment or attempting suicide in response to command hallucinations (Shawyer et al., 2003). Furthermore, positive symptoms, and lack of insight into these, can impact upon the therapeutic alliance between therapists and patients (Wittorf et al., 2009), which could stymie efforts to address cognitive symptoms, social function, and basic needs such as shelter or routine medical care. These symptoms can also have a significant effect on a patient's view of themselves in relation to the world. For example, patients with persecutory beliefs experience significant distress when entering crowded social environments (Freeman et al, 2015), which might lead to an increase in self-isolating behavior. Delusions of control might result in the distressing sense that one is not in total control of their own thoughts or actions, or that their autonomy is under assault by some outside force (Fierro et al., 2018).

Most existing pharmacological treatments are designed to help reduce positive symptoms, primarily via the blockade of dopamine in the mesolimbic pathway, though especially among the atypical antipsychotics a number of other neurotransmitter pathways are affected by these medications (Brisch et al., 2014). Despite years of research and drug development, however, many patients remain resistant to antipsychotic treatment (Potkin et al., 2020). A major focus of recent research has been better understanding the pathophysiology of psychotic syndromes in order to design new treatments (Stepnicki et al., 2018).

In addition, for many patients there exist a number of barriers to adherence to medication which may include: positive symptoms (Haddad et al., 2014) and positive patient attitudes towards positive symptoms (Moritz et al., 2013), poor insight (Higashi et al., 2013), negative symptoms (Haddad et al., 2014), cognitive symptoms limiting ability to obtain medications or adherence to treatment (Haddad et al., 2014), medication side effects (El-Mallakh & Findlay, 2015), or inability to afford medications, medical care, housing, or services (El-Mallakh & Findlay, 2015). Given these complexities, as well as the host of other social and medical challenges which face patients who live with psychosis (see (Lambert et al., 2003; Cohen 1993) for review), it is no surprise that for many patients the management of positive symptoms is not a simple matter of visiting a general practitioner, picking up a prescription, and taking their medications independently at home.

To address these difficulties, the delivery of treatment to patients living with psychosis has been marked in the last few decades by the growth in use and popularity of specialized, evidence-based clinical services, such as assertive community treatment (ACT) teams (Phillips et al., 2001) and coordinated specialty care (CSC) clinics for early psychosis (Kane et al., 2016). Far more than just mechanisms for the improvement of medication adherence, these services usually provide a number of "wrap-around" therapeutic benefits such as supportive and other psychotherapies, social worker support, support with ensuring access to medication, transport, and housing, vocational and educational support, coordination of care with other services, and regular followup, often occurring, in the case of ACT, in the patient's home (Phillips et al., 2001; Mueser et al., 2015).

The conventional view of these programs is that they work by providing the above services in an integrated manner, ensuring that patients get access to evidence-based treatments and services that support recovery and engagement in treatment. ACT teams have indeed been shown to reduce hospitalizations (Phillips et al., 2001), suggesting the services they provide may modify disease course. Furthermore, CSCs may help get patients comprehensive services during a 'window of opportunity' which exists in which services may help improve patient outcomes after first episode psychosis (Srihari et al., 2014). However, despite the clear and demonstrated superior outcomes and feasibility of both types of service (Phillips et al., 2001; Kane et al., 2016), there does not seem to be a clear *mechanistic* account of how these services produce their effect--that is, while it is clear that providing integrated services improves outcomes, *why* delivering services in this manner for these particular patient populations produces this effect remains an open question. This may seem trivial, in the sense that it seems obvious that providing personalized, integrated services would clearly be superior to providing one-size-fits-all, difficult to access treatment. However, as we will argue below, examining how these services are provided, and how that might interact with a hypothesized computational model of the experience of psychosis, may reveal a number of interesting possibilities for improvement of services and novel mechanistic considerations.

ACT teams and CSCs have been shown to help reduce symptom severity more than standard treatment (Craig et al., 2007; Nossel et al., 2018). While these accounts certainly explain why a patient's quality of life would improve while receiving these services--for example, a patient receiving ACT support gains a team that can help them manage their household and gain access to needed medical care and social support--they do not directly explain why a patient would have reduced symptom *intensity*. One possibility is that these programs assist patients in managing their lives in a manner that reduces the occurrence or impact of stressful life events, which have been shown to be significantly associated with re-hospitalization or relapse of psychotic symptoms (Martland et al., 2020); if this is true, one can ask further how we might characterize the interactions between patient and team which lead to stress reduction. Another possibility is that patients are simply taking their medications more regularly. As noted above, medication may be a necessary but in many cases insufficient part of the control of positive symptoms. In addition, Guo et al. (2010) found that medication plus psychosocial support outperformed medication alone in early schizophrenia, though this result is confounded by the fact that treatment discontinuation rates were higher in the medication-only group. Perhaps the key point to be made is that, despite ACT and CSC teams varying in composition and services provided (Philips et al., 2001; Srihari et al., 2014), none of them are limited to services that simply improve medication adherence. As such, considering all of the limited evidence available, one may hypothesize that other services in addition to the provision of medication, and potentially the experience of these services, may have an important impact on symptoms. Thus, we return to the question of mechanism: how do these external programs have an impact on the internal experience of a person, and of their experience of their self in the world and in relation to their symptoms?

Examining this question requires engaging in one of the most challenging exercises in behavioral science- integrating information across levels of explanation (Boccignone & Cordeschi, 2015), or attempting to understand how a change in social or psychological situation might affect neurobiology or vice versa. In attempting to integrate across these different levels of interpretation and analysis, it is helpful to have one common language- a set of concepts which can serve as the benchmarks for use in comparing pieces of information from different levels of inquiry. A key requirement of this common language would be that it can assist in the understanding of how the patient's self interacts with both the external world and with positive symptoms (like hallucinations) that are often experienced as being separate from them. One such language (or perhaps more accurately, set of languages) is that of computational psychiatry (Huys et al., 2016), a discipline that seeks to understand what, precisely, the brain is trying to compute, how it implements this computation (from both a psychological and neurobiological standpoint), how these processes influence and are influenced by the environment, and how psychiatric symptoms may arise from aberrations in any number of these processes. Once one has an idea of *what* the brain is trying to compute, one can explore the possible ways in which the brain might implement these computations neurobiologically; one can then observe how the quantity being computed changes with a change in the environment, and then deduce what changes at the level of neurobiological implementation would be necessary to allow these changes computation to obtain. Crucially, one must be able to use this model to gain an understanding of how to tie neurobiologically implemented computations to socially-oriented clinical interactions, which in turn may shed light on the way the patient's self interacts with external actors and with their symptoms..

A number of computational theories of psychosis have been proposed (Sterzer et al., 2018, Adams et al., 2013). In this conceptual paper, we will select one particular theory- the active inference account (Benrimoh et al., 2018). As we will discuss, we have selected this theory because of its enactive nature, its existing physiological validation, and the manner in which it explicitly considers the role of an agent in the context of its environment. However, the choice of theory is not the main focus of the paper, and another theory could have been chosen. Rather, the focus of the paper is to demonstrate how, using a computational theory as a framework, we can conceptualize a mechanism of action for the specialized services described above and relate this directly to potential underlying neurobiology. We will then demonstrate how experiments could be designed to test the veracity of this hypothetical mechanism of action.

BACKGROUND ON ACTIVE INFERENCE AND OTHER COMPUTATIONAL ACCOUNTS OF ACT/CSC

In this concept paper, we have chosen to focus on the active inference theory of hallucinations in order to provide the computational language we will use to investigate the potential mechanism of action of services such as ACT teams and CSCs. We begin with a brief overview of active inference. Though more substantive reviews and descriptions of its mathematical underpinnings are available (Friston et al., 2017), we instead focus here on describing active inference in a manner that will be accessible to clinicians and clinical researchers.

Active inference is a Bayesian theory (Friston et al., 2017) which, at its core, asserts that all agents (organisms acting in the world) function in a manner that aims to minimize the uncertainty in their model of the world. This uncertainty can be quantified and is often referred to as surprise or free energy (Friston, 2020); decreases in free energy can be projected in an approximate form, and this in turn can guide the selection of action (i.e., agents take actions which help reduce their uncertainty). The theory is one of 'inference' because agents do not have access to the processes in the outer (or inner) world that cause their sensations- they must infer these causes. It is 'active' because agents must use actions in order to gain new information to test their models or update them. It is Bayesian because the agent has a set of prior beliefs about the world, takes in new observations (often after acting in order to acquire said observations), and then forms new beliefs (known in Bayesian theory as posteriors). Any action--or series of actions--that an agent can take is called a *policy*; the set of actions or series of actions that an agent can take is called a *policy space.*

As a simple example, imagine you are in a dark room. You do not know anything about the room or what may be contained in it. You are, however, armed with certain prior beliefs about yourself and the way rooms are generally constructed. For example, you believe that you have a pair of eyes that require an external light source in order to see and you also believe that most rooms are equipped with a light switch. You look around and notice a pair of eyes shining in the dark. You are now in possession of a sensory observation--a pair of shining eyes--but you do not know what process in the world has *caused* that sensory observation- i.e. to whom or what the eyes in question belong. You have several policies available to you in your policy space--for example, you could walk towards the eyes; you could stay still; you could try and find the light. Of all of these actions, turning on the light--you project--has the highest chance of reducing your uncertainty the most, and as such you select this as your policy. You proceed to find the light switch and discover that the pair of eyes belonged to a tabby cat. You then update your model of the world- the sensory observation that was the eyes was likely *caused* by the cat (the presence of which you again infer from your sensory observation once the light is turned on). The belief that there is a cat in the world that can cause the sensation of shining eyes in the dark becomes a posterior belief; it can at a future time serve as a *prior* belief. For example, if you choose to turn the light off you might *expect* to see the shining eyes in the dark again- all due to your now-prior belief about the cat and the sensations it can produce. It is also key to understand that *policies can engender expectations about what is to be experienced.* For example, if you turn on the light you *expect* that, in the next time step, you will experience the light filling the room and rendering objects more easy to see. In this way, policies can act as models of how one might expect our perceptions (the causes of which we must infer) to evolve over time, in response to the actions- or inaction- that one might take. In other words, policies can serve as models of how we expect the world to evolve, and in turn a policy space defines the set of possible evolutions we may consider.

There is one more key concept to understand in our simplified description of active inference- that of *precisions*. In the real world, not all beliefs are held with equal certainty, and not all sources of information are equally reliable. For example, if you had poor eyesight in our example above, then when the light was turned on you might be only partially certain you are seeing a cat, which in turn might influence what you end up inferring to be the cause of the shining eyes in the dark. Equally, depending on your experience with usual room construction you may have more or less certainty in the different policies available to you while you are in the dark room (i.e. if you had been transplanted from a time just before electric light switches became commonplace, you may only be somewhat certain that searching the walls of the room might reveal a switch for an electric light). These varying levels of certainty in observations, beliefs or potential courses of action are called *precisions* under active inference and act as coefficients in active inference equations which can be manipulated in order to alter agent behavior.

Armed with an understanding of the basic aims and computational elements present in an active inference agent, let us now turn to a discussion of how one might cause such an agent to experience positive symptoms--for example, hallucinations. In (Benrimoh et al., 2018) we describe several situations in which this can occur. Let us imagine an agent who must infer its current state- whether it is listening to someone speaking or whether there is silence. In terms of actions it is able to speak or to listen, and it must select between a set of policies which will have it listening or speaking at different points in time. It is paired with another agent that speaks every second time-step (an arbitrary choice in this case, chosen for simplicity), and which serves as its external 'world' or environment. Finally, our agent has two different precisions to consider. The first is *sensory* precision. This is the confidence it places in information it observes- for example, auditory information. This confidence might be a result of the integrity of the information- dipping, for example, in the case of external noise or poor transmission over internal networks, such as the reduced integrity of the arcuate facilsulus seen in schizophrenia (Catani et al., 2011); it might also be a result of how much attention- which can be formulated under active inference in terms of sensory precision- is being placed on the information (a reflection of its perceived utility), which in turn has been linked to acetylcholine signalling (Moran et al., 2013). The second is *prior precision over policies,* or the strength of the agent's belief in the policies in its own policy space; this certainty about actions (or lack thereof) has been linked to dopamine signalling in previous work (Schwartenbeck et al., 2015).

      The first way to make this agent hallucinate is to drop its sensory precision. In this situation, where it cannot use information from the environment to correct its prior beliefs, then these priors become dominant: when the agent expects to be listening to the other agent in the world (as a result of the policies it has chosen to pursue, which involve listening), it can *infer* that sound is present because it expects it to be- in other words, it generates the percept of sound (a hallucination) in order satisfy its prior belief in the absence of information that could be used to correct this belief. If one were next to lower the sensory precision, but not to the same degree as previously, in an agent with an elevated prior precision over policies, the agent would similarly hallucinate. In this case, the increase in certainty about priors tips the balance, leading to hallucinations when observations in the world are not reliable enough to counteract these priors. Let us examine the policy space more closely. It is possible, as we have demonstrated (Benrimoh et al., 2018), to construct policy spaces that never result in hallucinations- in this case, because they happen to match what happens in the world perfectly. These agents do not hallucinate even if sensory precision is reduced significantly, because their expectations and what the world produces match up. This is a very fragile state, however: when the policy that matched the world is 'lesioned' and removed from play, the agent is once again vulnerable to hallucination. In other words: when an agent has a healthy and expansive policy space and a normal ability to use information in the world to correct faulty prior beliefs, then hallucinations do not result. As an example, imagine an agent playing a chess game and possessing a given moveset. The agent can use any of their moves to win the game and to keep up with the environment in the game changing (as a result of the moves of the other player). If we shrink the agent's moveset and allow it only one set of moves that is consistent with winning the game, the agent can still perform well but things are fragile- change the environment (i.e. increase the unpredictability of the opponent, for example), or further decrease the moveset, and the agent will no longer be able to produce moves consistent with a valid solution to the game. In our case, the "chess game" is perception and interaction with external reality, and incorrect solutions to that game driven by a limited moveset would be hallucinations or delusions. We should also note that, in addition to priors provided by policies, maladaptive perceptual priors can also be learned over time- perhaps as a result of continuous experience of maladaptive policies; these perceptual priors could then take the place of the main prior driving generation of hallucinations (Powers et al., 2017). Reduce the functional space of policies the agent can exploit- creating an agent maladapted to the world-, increase this agent's certainty about maladaptive priors, and, finally, reduce its ability to correct these prior beliefs by referring to external information and you create an agent that hallucinates because its prior beliefs have gained dominance over the evidence it gains from the world. A further elaboration of this model which expanded it to the *content* of hallucinations was described in (Benrimoh et al. 2019). Here

the agent is singing a song with another agent, and may believe the song order is standard or altered (i.e. the words of the song have been moved around). It will only have hallucinations of the song out of order when it believes that the song is disordered *and* its sensory precision is low, robbing it of the ability to use the other agent's singing to correct its faulty prior beliefs. Agents with abnormal prior beliefs about word order in the song who are still able to use external information- i.e. whose sensory precision is intact- do not hallucinate.

To summarize, one can conceptualize hallucinations- and perhaps delusions- as arising from an imbalance between prior beliefs- for example, beliefs that one should be hearing someone speaking, and that the speech will include certain content- and sensory information, where priors have won out over sensory information. There has in fact been recent experimental evidence for this 'strong priors' account of hallucinations (e.g. Vercammen et al., 2010; Teufel et al., 2015; Corlett et al., 2019); though as we discuss in (Benrimoh & Friston, 2020), the dialectic between 'strong' or 'weak' priors is likely one that can be dissolved by turning instead to a discussion of hierarchy and timing within the disease course.

The key reason for using this computational framework is that it allows for agents to actively predict the world around them and to act in that world and *interact* with it. Active inference can be used to model social interactions- indeed the above models explicitly describe the generation of hallucinations in a social context, where another agent is present in the world and which the hallucinating agent can interact with and try to predict. This helps us understand how a patient may interact both with their generated hallucinations- which can be experienced as an external agent- and with external actors, who can provide information that may be of use to the patient in helping them understand- and predict- the world. In addition, and as noted above, it provides a framework that can link this social level to the neurobiological level. We will now turn to a discussion of how this conceptualization might assist us in probing the underlying neurobiology before then ascending the levels of interpretation and returning to the question of the mechanism of action of programs such as ACT or CSCs.

GENERATIVE MODELS AND MECHANISMS OF PSYCHOSIS

Psychiatric treatment has long depended on models of symptom development: Freud's structural model proposed that psychiatric symptoms could result from unresolved conflicts between biological drives and internalized social limitations (Freud, 1923); behaviorist and cognitive models attempted to ground psychiatric nosology in emerging knowledge of learning, memory, and information processing theory (Neisser, 1967). The advent of biological psychiatry brought with it (simplistic) brain-based explanations of the roles of different neurotransmitter systems in the generation of basic psychiatric symptoms, often arising as post-hoc justifications for the use of effective treatments discovered by serendipity (e.g., the monoamine hypothesis of depression (Delgado, 2000), the dopaminergic hypothesis of schizophrenia (Meltzer & Stahl, 1976)). Each of these models drove significant advances in the recognition and treatment of psychiatric illness. However, these advances have fallen far short of the standards set by the rest of medical nosology, which uses knowledge of pathophysiology to inform objective tests to identify specific disease states and predict prognosis and response to treatment.

Computationally-informed generative models attempt to establish how latent information processing states give rise to symptoms. Using formal mathematical systems to describe how information is used to produce behavior provides several tangible benefits. First, formalized systems tie hypotheses to measurable quantities derived from behavior, eschewing vaguely-posited theories of symptom development in favor of quantitative variables. This means that formal generative models are inherently falsifiable. This also means that experimental work can be applied more quickly to the development of actionable tests and biomarkers, based as they are in measured, replicable quantities derived from behavior and other objective data. Additionally, because the most useful models are capable of identifying and then discriminating among different latent states that drive behavior and may differ across groups, they are in principle also capable of identifying meaningful subgroups of individuals exhibiting the same symptoms for different mechanistic reasons. This is the equivalent of identifying individuals with a symptom profile consistent with fluid overload (i.e., dropsy (Clementz et al., 2015)) who have underlying heart failure, kidney failure, cirrhosis, or blood dyscrasias: by identifying differential causal pathways, clinicians might intervene differently to arrest or reverse the causes giving rise to clinically indistinguishable symptoms.

The ability to intervene depends on a generative model that is sufficiently rich and sufficiently tied to a plausible biological instantiation in the brain. The Bayesian family of generative models (including Active Inference models) is particularly well fleshed-out in this respect (Friston et al., 2014; Friston 2010). Similarly to the work above describing how computational parameters have been tied to neurotransmitter systems, recent data have brought evidence to some of these ideas and tied symptomatology to specific neurotransmitter and circuit abnormalities via computational parameters (Powers et al., 2017; Cassidy et al, 2018; Alderson-Day et al., 2017, Howes et al., 2020). For example, Powers et al., 2017 directly correlated parameters from computational model fit to fMRI data demonstrating hyer-precise perceptual priors in hallucinators. Cassidy et al., 2018, linked striatal dopamine release to the precision of predictions and resulting hallucinations.

Within this framework, perception may be thought of as the process of inferring the causes of one's sensory input given existing knowledge about those causes as well as the input itself, weighted by the reliability of these sources (Adams et al., 2013; Friston, 2005; Powers et al., 2016, Corlett et al., 2019). An extension of these ideas into the realm of active inference is outlined above. By couching the emergence of hallucinations within these formal terms and in the context of a hierarchical system of brain organization, we are able to make predictions as to how interventions as diverse as social support and brain stimulation may act in concert to correct or compensate for alterations leading to dysfunction in hallucinations. Here, we propose that the interventions shown to be effective in specialty care for psychosis may exert that efficacy in predictable and consistent ways within the AI framework.

In the following section, we examine the existing evidence for the mechanisms underlying effective components of specialty care for psychosis and then turn toward the integration of this evidence with other elements of the AI framework.

PROPOSED MECHANISMS OF SPECIALTY CARE

Before turning to a discussion of how an AI-based understanding of positive symptoms and its putative neurobiological correlates may be related to the efficacy of specialty interventions for psychosisACT/CSCs, it would be helpful to discuss how these interventions may exert their desired effects. It is worth noting that, while the efficacy of these programs is well documented, there is a lack of research postulating the mechanisms that underlie the programs' success. Of the literature that does address mechanism, few papers propose a specific model or approach for investigation, much less a computational one. As discussed in (Bosanac et al., 2010), one limitation of existing research is that interventions are multifaceted, making it difficult to disentangle the elements within the intervention in order to create causal models. It should also be noted that most papers examining the effectiveness of ACT and CSCs show efficacy in terms of global symptomatology, hospital readmission, and social and occupational function, but not in terms of specific symptoms or symptom clusters (such as auditory hallucinations or positive symptoms distinct from negative symptoms)(Ziguras and Stuart 2000; Vijverberg et al. 2017).

In a recent study from Daley et al. (2020), which analyzed semi-structured interviews from individuals enrolled in CSC programs across the United States, a majority of participants reported what they perceived as the mechanism through which change had occurred during their time in the program. Participants pointed to therapy, medication support, and instrumental support (case management, employment and education services, etc). They also noted the importance of forming meaningful relationships with their providers (Daley et al., 2020). Some other frameworks that may be useful for interpreting these results include the Therapeutic Contracting Model (Heinssen et al., 1995) and the Health Belief Model (Fenton, Blyler, & Heinssen, 1997), which view patient-centered treatment, perceived physician investment, treatment efficacy, and environmental interventions as important drivers of favorable health behavior. The focus on patient-centeredness and shared decision making may also be important in the improvement of medication adherence in this population (Deegan et al, 2006). While these frameworks provide initial insight into the components of treatment that contribute to the success of CSCs and ACT, our paper offers a model for understanding at a deeper level why these reported cornerstones of care may be effective in reducing symptoms.

PROPOSED THEORETICAL FRAMEWORK

We now turn to our proposed theoretical framework for applying the computational language of active inference to ACT teams and CSC's in order to understand potential mechanisms of action for these services.

Let us begin with ACT teams. This particular service is often employed to support patients with more chronic forms of psychotic illness, especially when these patients are isolated from family or other social contacts (Pettersen et al., 2014). Under the above active inference account, these patients are in a state where the precision of their observations is reduced--that is, the confidence in, and therefore the attention afforded to, observations that may oppose their hallucination- and delusion- inducing priors is reduced. This is concordant with evidence that deficits in attention and other cognitive domains is a feature of schizophrenia that can impair social and occupational functioning (Harris et al., 2007; Green, 1996; Mucci et al., 2021). Two solutions to this situation seem plausible. One would be to reverse the tendency to afford low precision to external information--or in other words how external information is filtered--via some change *within* the patient, as might be produced via pharmacological treatment or after a course of cognitive training or CBT for psychosis (Harris et al., 2007; Bell et al., 2009; Morrison, 2009) For example, a therapist might help a patient to re-evaluate their persecutory narratives and the trustworthiness of other people (e.g. family or members of their medical team), or cognitive training may help patients better attend to the environment around them in order to perform important tasks of daily living. Another approach would be to provide external information of greater perceived reliability which might encourage the patient to, over time, increase the precision afforded to information. These approaches are not mutually exclusive (and, in fact, may reinforce each other), and both seem to be present in the ACT team approach. The ACT team provides reliable, sometimes daily, social contact with professionals who provide consistent information about treatment as well as socioeconomic support. The fact that this support and information is delivered by other people in a consistent manner may be critical, given that patients with schizophrenia are less sensitive to social cues (Corrigan & Green, 1993) and thus may require more consistent social engagement; the 'assertive' nature of ACT also defeats the potential for patients to use social withdrawal to avoid this source of information. In addition, the provision of supportive psychotherapy--or even of CBT for psychosis interventions--as well as the structured nature of the ACT program itself may help to gradually re-train the patient's attention in order to improve the precision afforded to external sensory information. In this context it is interesting to consider the experience of someone experiencing ACT services: one might imagine the experience of psychosis when one is isolated to be similar to that of an echochamber- one in which one's own priors dominate experience, where the self effectively dominates external perception because of low sensory precision afforded to external cues. ACT services provide close social support, and regular contact with an external world that *makes* itself salient because of the assertive nature of ACT as described, and which provides a way out of the "echochamber" of priors run rampant. Indeed, in a study of patient experiences of ACT, one of the positive patient experiences was related to increased social contact and decreased isolation (Watts & Priebe, 2002).

      The repeatedly demonstrated effectiveness of ACT teams in preventing rehospitalization and relapse may have significance with respect to computational patterns as well. If the maladaptive priors that we argue underlie at least some psychotic experiences are reinforced when they are experienced in full (i.e., during relapse), then reducing the frequency these symptoms are experienced may, in some individuals, begin to shift patient priors away from the maladaptive set. The idea that psychotic symptoms are reinforced during relapse is supported by the literature demonstrating poorer outcomes correlating with greater relapse rates and rapid onset of symptoms during relapse (see Emsley et al., 2013, for a discussion); this is the behavior one would expect if psychosis is in fact an attractor state (Adams et al., 2013) which is 'deepened' and made easier to fall into with repeated exposure; in addition, one might expect the time required to recover from this state to lengthen after each subsequent relapse, which is true for some patients (Emsley et al., 2013). Reducing the experience of psychotic symptoms may therefore interfere with this process or provide alternatives that can weaken the kinds of perceptual priors described in Powers et al., 2017. In addition, the utility of ACT teams in improving medication adherence (Manuel et al., 2011; Dixon et al., 1997) provides another mechanism for the reduction in positive symptoms.

      It is interesting to consider, however, what happens when ACT team support is withdrawn, as this may help us understand to what extent ACT provides permanent versus temporary change in patient capacities. While many ACT teams try to set rehabilitation goals for their patients, a number of patients require sustained services by ACT or, once they transition from ACT, continued community services, and in one study more than 10% of patients experienced psychiatric rehospitalization within 90 days of the end of ACT services (Huz et al., 2017). Why might this be? Thinking back again to our computational model, there are several possibilities. Firstly, the reduction in medication adherence in the absence of the ACT team's close follow-up may lead, via increase in dopaminergic activity, to a re-emergence of strong priors over a maladaptive policy space, making the mind of the patient more vulnerable to psychosis. Secondly, the lack of the ACT team's role in enriching the reliability of information in the world may result in a return to low sensory precision incapable of counterbalancing strong priors, the prerequisite combination for hallucinations. If patients have more severe illness, then it is also possible that the attractor state of psychosis has already become too deep, and the shifting of maladaptive priors toward more adaptive priors described in the previous paragraph may not be possible; this may explain why some patients require chronic ACT or similar-intensity services. Let us turn to the other key element of the computational model: the policy space.

The policy space represents the set of actions or sequences of actions a patient may take in the world, and represents the potential for the patient to adapt to changing circumstances. When their policy space is limited and does not accurately account for the environment, and the information from the environment has low precision, then hallucinations can occur. These are driven either by sensory priors formed over time, or by the prior expectations of sensory information driven by policies which affect state inferences. As above, if sound is *expected* (as a result of a sensory prior) or one *expects to be listening to a sound* (as a result of a maladaptive policy), and no high-precision information is available to sway perception towards the correct state (the inference of silence in the world) then sound will be *generated* in the form of a hallucination in order to satisfy the prior. How might an ACT team interface with the policy space? Two possibilities seem evident. The first is that the ACT team, through role modelling or the provision of therapy, help to *expand* a patient's policy space through learning. The second is that the team assists directly in problem solving- such that when the world is not explained by the set of policies the patient has at their disposal, the team can assist the patient by effectively 'adding' their policy space to the patient's. The patient would still need to grow their policy space sufficiently to include the policy of "working with the ACT team", but this may be a less taxing requirement than learning specific policies to deal with each of the other challenges life can generate. As such, when ACT team services are withdrawn, some patients-- those for whom it is still possible to grow their policy space, potentially because of less advanced or severe illness--may still demonstrate improved function because of what they have learned while they were receiving services, whereas other patients--those who were relying on the ACT team as an external cognitive assistance mechanism--would be more likely to respond to new or stressful events, for which they lack the appropriate policies to respond, with a relapse of positive symptoms. This phenomenon is certainly observed clinically: many patients with chronic schizophrenia are re-hospitalized after stressors (such as altercations with landlords or interactions with government authorities) for which the response is recurrence or exacerbation of psychotic symptoms (Martland et al., 2020).

Thus far, we have discussed how ACT teams might improve the functioning of patients with chronic psychosis in three main ways, using the language of active inference and computational psychiatry. Firstly, ACT teams provide a stable and reliable source of information and social contact, which may enrich the information in the outside world available to a patient or which might help them learn to better use the information already available (increasing sensory precision). Secondly, the teams ensure access to and often adherence to medication, which acts by reducing dopaminergic signalling (reducing prior precision over policies). Finally, the teams may help patients learn how to approach life challenges more effectively (growing the policy space) or at least to help patients problem solve (by acting as a 'supplemental' policy space). These mechanisms may be difficult to separate out, and likely work in concert with each other, but *which* particular mechanism is more crucial for a given patient, based on their individual capacities, may be relevant in determining the type, level, and longevity of services required, and the response of the patient to the withdrawal or interruption of these services. This latter point about service interruption is relevant because services are commonly interrupted- staff may go on vacation, maternity leave, or a holiday may reduce the number of available staff. Having a better idea of which patients are at highest risk for relapse in the event of temporarily limited services may assist in the assigning of temporary or replacement services to patients in greatest need.

We turn now to a discussion of CSCs, which primarily serve first-episode patients and serve as the basis for approaches meant to treat those at Clinical High Risk for psychosis

(CHR-P). In addition to symptomatic relief and assisting in ensuring function, many of these clinics also have as a mandate either the early detection and treatment of psychosis in order to improve symptomatic and functional outcomes (Albert et al., 2019). While research is still ongoing to determine long-term outcomes in patients treated in CSC clinics (Chan et al., 2019), it may still be of some utility to consider their possible mechanisms of action for the fulfillment of these two mandates from a computational perspective. As discussed in (Benrimoh et al 2019), the prodromal or high-risk state can be conceptualized as one in which there has been some change in the priors a patient may hold- beliefs may have begun to grow more anxious or fearful (for example, a person may develop beliefs of being persecuted or of being watched), prior to and then alongside the development of attenuated psychotic symptoms (consistent with the description reviewed in Powers et al., 2020). However, there seems to still be enough ability for patients to be able to use the context present in the environment to counteract these maladaptive beliefs and to avoid frank psychotic symptoms. The reasons for this preserved ability are relevant.

One possibility is that these patients retain at least moderate sensory precision. One reason may be because the neural pathways that encode this precision have not yet lost their integrity. Previous work has demonstrated interval deterioration in gray matter and white matter tracts between the high-risk state and first-episode psychosis (Jung et al., 2011; Ziermans et al., 2012), suggesting that ongoing neural deterioration may be leading to changes in the quality of information carried over time. Another possibility is that increasingly maladaptive and rigid priors over the course of the progression of the prodrome lead patients to actively *reduce* their sensory precision. Normally, a prior shown to be incorrect based on gathered data would be updated. If this prior is rigidly fixed- for example, because of a reduced or over-confident policy space which does not allow for other alternatives- then a dilemma occurs: should the agent trust its priors or its observations? If the prior cannot be changed, then a reasonable course of action would be to begin to doubt incoming sensory information- to reduce its precision. This may occur, as noted in the previous section, via a modulation of ACh. Early intervention while there is still some flexibility in priors or before sensory information has been fully down-weighted may delay or even prevent the onset of psychosis, although the type of intervention chosen would depend crucially on the pathophysiological processes at play in hallucinogenesis. Importantly, careful identification of potential computational subgroups within these early-stage patients may be crucial for the development and personalization of interventions. For example, for some patients with low sensory precision because of reduced ACh but intact neural pathways, increasing sensory precision may be beneficial; in other patients with less preserved neural pathways, increasing the precision of incoming information may be less effective as it might result in the amplification of inherently noisy signals. What remains to be determined, however, is the relative importance of neuro-degenerative processes versus altered neural signalling in the mechanisms underlying the transition to psychosis. The former may be less amenable than the latter to purely socio-environmental interventions, though we hypothesize that the manner in which neurodegeneration proceeds (i.e. those policies which are pruned versus those which are preserved) may depend on the prevailing emotions and beliefs held more or less strongly by the patient during the process. These in turn may be amenable to psychotherapy and social interventions.

Another possibility is that FEP or high-risk clinics can help patients navigate the very challenging situations which can occur during the FEP or high-risk period- such as difficulties in scholastic or professional life caused by symptoms, and the transition from a state of wellness to one of relative disability. This is consistent with the focus of CSCs on case management and an individualized approach that adapts to the needs- and specific stressors- experienced by each patient. They may also help patients recognize, explore and understand emerging psychotic beliefs and experiences while maintaining and reinforcing links to consensual reality, preventing them from developing the crystallized set of delusional beliefs that in commonly used scales such as the PANSS characterize more severe illness (Kay et al., 1989). This rigid set of beliefs may inform the maladaptive policy space that we hypothesize is present in more chronic illness, and any efforts that could be made to retain flexibility or include more adaptive policies may help to prevent or reduce the severity of further illness. Additionally, early intervention may prevent or mitigate the formation of maladaptive perceptual priors- the deepening of attractor states described above- which coheres with the philosophy of CSCs that early intervention and decrease in the duration of untreated psychosis is key for improving long-term outcomes. Overall, concurrent with the high-risk and FEP literature, there appears to be a computational argument for concerted attempts at prevention, treatment, and mitigation of psychotic symptoms at this phase of illness. From an experiential perspective, the thesis here is that less experience of psychotic symptoms would prevent the need to develop the crystalized, internally consistent psychotic narrative which characterizes more severe illness- keeping the person more grounded in consensual reality from the start, rather than needing to bring them back to that baseline from the thoroughly different experience of repeated psychotic episodes.

PROPOSED STEPS FOR VALIDATION

In order for these models to be of use in assisting with the improvement or innovation of treatment strategies, there must be the possibility to test the hypothetical mechanisms provided and to either support the proposed mechanisms, to falsify them, or to suggest alterations that may improve them. Given the complexity of the cognitive and emotional processes at play and the multiple processes which may give rise to observed behavior, designing experiments to test specific cognitive mechanisms can be challenging, though recent work has demonstrated that it is possible (Vercammen et al., 2010; Teufel et al., 2015; Powers et al., 2017; Benrimoh and Friston, 2020). We will discuss some options here.

One possibility is to conduct an imaging study of patients in ACT followup and to relate white or grey matter integrity to the length of required ACT followup. This would support the conceptual model if patients who have a greater burden of neural degeneration- which might relate to inflexibility of priors or policy spaces- require prolonged care from ACT services as these services help provide the external computational resources described above. However, this would not provide very specific support to the model, as lower neural integrity may simply reflect more severe disease overall. To provide more specific support, we could also try to determine if this relationship between reduced neural integrity and need for ACT support is moderated by measures relating to our computational parameters. These could include tasks that measure strength of perceptual priors (e.g. Vercammen et al., 2010; Teufel et al., 2015; Powers et al, 2017) or new tasks, currently in development, which measure policy space size and flexibility. These may take the form of open-ended games which patients may be able to solve in a number of ways, with the number of solutions utilized being a proxy for the size of the policy space. It would also be important to correlate these tests with measures of the cognitive deficits often seen in schizophrenia (Bowie et al., 2006) in order to understand how they may be related and whether more general cognitive training strategies may improve this measure. These psychophysical tasks could be further correlated with need for specific ACT services, such as problem solving support. A lack of correlation between performance in these tasks and need for ACT services would serve- importantly- as strong evidence against the conceptual view presented here. An expansion or maintenance of policy space after ACT or CSC services compared to treatment as usual would also provide evidence in favor of this model, as would a reduction in the strength of perceptual priors after ACT or CSC services. It should be noted that the possibility for the expansion of policy space or reduction of the strength of perceptual priors may depend on disease severity; as such, maintenance of existing policy space with ACT compared to further contraction with relapse in treatment as usual may be expected for more severe patients, and the lack of strengthening of perceptual priors with CSC compared to treatment as usual over the course of disease progression may be expected for patients with more severe first episode psychosis.

With respect to testing the neurobiological implementation of this conceptual model, one can imagine a number of experiments. For example, to test the role of ACh in encoding sensory precision, tracer studies could look at changes in ACh signalling before and after initiation of ACT services, in order to determine if ACh increases with the onset and maintenance of ACT services. Similarly, to test the role of decreased sensory precision in the onset of psychotic symptoms, CHR-P patients could have ACh signalling serially tested before and after a first episode psychosis, to determine if a reduction of ACh correlates with psychosis onset and if treatment at a CSC is correlated with ACh increase alongside reduction in psychotic symptoms; a comparison group treated outside of a CSC could also be included to determine if CSC services have an effect beyond treatment as usual, which mostly centers on dopamine-blocking medication without the extensive services provide by CSCs (an important caveat to strict interpretation of these experiments, however, will be the interactions between the dopaminergic and cholinergic systems). In addition, given recent positive preliminary studies of ACh-agonist agents in reducing symptoms in schizophrenia (Brannan et al., 2021), one might also consider testing if these agents might help to arrest symptom development in at least some sub-groups of patients in prodromal or early phases of psychosis. Once again, psychophysical measurements of perceptual priors, sensory precision, and policy space could also be administered at different time points in order to determine how they correlate with both service use, symptoms, and neurobiological measures. These psychophysical and neurobiological measures, administered in high-risk groups served by CSCs, may also have value in predicting transition to psychosis, as has been the case for related measures, such as the repetition positivity (Fryer et al., 2020) or mismatch negativity (Perez et al., 2014) measured by EEG.

Finally, qualitative research should be a significant component of the above experiments, as this would help to 'close the loop', so to speak, and to related these measures, measured symptoms, and the experience and views of people with respect to their hallucinations and their relationships with caregiving services like ACT teams and CSCs.

POTENTIAL SIGNIFICANCE

At first glance, one might argue that adding the computational lens does not add very much to the account of mechanisms of action of ACT teams and CSCs. Using this language has perhaps provided a sophisticated theory for why social or problem solving support may benefit patients, but it has not changed the fact that these well-established interventions are effective. Indeed, the components of the interventions identified as being crucial here do not differ greatly from more traditional accounts relating to the reduction of stress or the provision of housing or social support discussed above. In addition, the computational mechanisms proposed are not easily separable, as several of them are affected by the same interventions, which seems to limit explanatory power. However, this computational language does provide us with two key advantages. The first is that it can be used to generate specific and testable hypotheses that can be used to improve existing interventions or potentially design new ones from first principles. The second is that it provides a bridge from a macro-level clinical intervention with multiple components, to potential psychological processes which could be measured with psychophysics experiments, all the way down to the potential neural implementation of these processes which can be interrogated using tracer studies and neuroimaging. This in turn allows for the synergistic design of interventional combinations, pairing psychosocial interventions with pharmacological interventions meant to optimize combined efficacy. At more social and experiential levels of interpretation, it also allows us to better understand the experience of patients in their interaction with services provided by CSCs and ACT teams. This paper serves as a case study for how thinking about mental disorders from the lens of computation allows us to integrate information across all levels of abstraction, and to use this information at different levels in order to design experiments or improve interventions.

We have laid out a number of possible experiments which might be used to establish (or falsify) some of the proposed computational mechanisms described here. These experiments, should they yield new metrics, may have clinical value. For example, should we successfully create tests that measure policy space or sensory precision, and demonstrate that they in fact behave as expected in this conceptualization, these could be used as tests for staging patients, determining progress, and attributing new resources as needed. Patients may also have care personalized to them based on their specific computational profile. For example, a patient with a limited policy space as a result of degeneration of cortical structures may be recognized as needing long-term problem-solving services, while another patient with a relatively preserved policy space may be enrolled in a program with a greater focus on rehabilitation- instead of subjecting both patients to the same rehabilitation-focused program and considering the first patient to have 'failed' treatment. New treatments- either pharmacological or environmental in nature- might also be designed with this computational language- and its neurobiological correlates- in mind, and with the use of tests derived from computational theory to act as biomarkers for treatment effects. Finally, researchers investigating other psychiatric conditions for which fewer treatments exist- for example, antisocial personality disorder- may use this case study as an example of how an understanding of underlying computations may drive our understanding of both neurobiology and potential interventions.

FIGURE LEGEND:
The following figures depict the perception of an agent given the differing precision of sensory priors, priors over policies, and size of policy space. An agent perceives a forest and may or may not hallucinate a singing bird (red bird). This incoming sensory evidence is either perceived clearly (crisp lines of the trees) or with noise (blurry trees with extraneous lines) in order to depict changes in sensory precision. Please note that these pictures are meant to provide a reader with an intuitive 'feel' for the concepts in question; they gloss over some key subtleties which are relevant to the mathematical understanding of these parameters.

The thought bubble shows two possible prior expectations about incoming sensory evidence, with bolder lines representing increased weight on that prior. The rectangle (policy space) contains strings of actions that the agent could take to gain more information about its environment: looking (eyes), looking closer with binoculars, listening (ear), moving closer (footsteps) Again, when these actions are represented with bolder lines, this denotes that they have more weight for the agent (i.e. that the agent is more certain it is engaging in the bolded policy).

Figure 1:

The perception of an agent that does not hallucinate is shown on the left. The incoming sensory evidence about the environment has a normal confidence ascribed to it and the agent has an expansive policy space without bias towards a specific string of actions. The agent does, however, have a strong prior belief that it will perceive a singing bird.

On the right, when precision of incoming sensory evidence is reduced and the forest scene looks blurry, the agent is unable to use the environment to correct for their strong prior belief and hallucinates the singing bird.

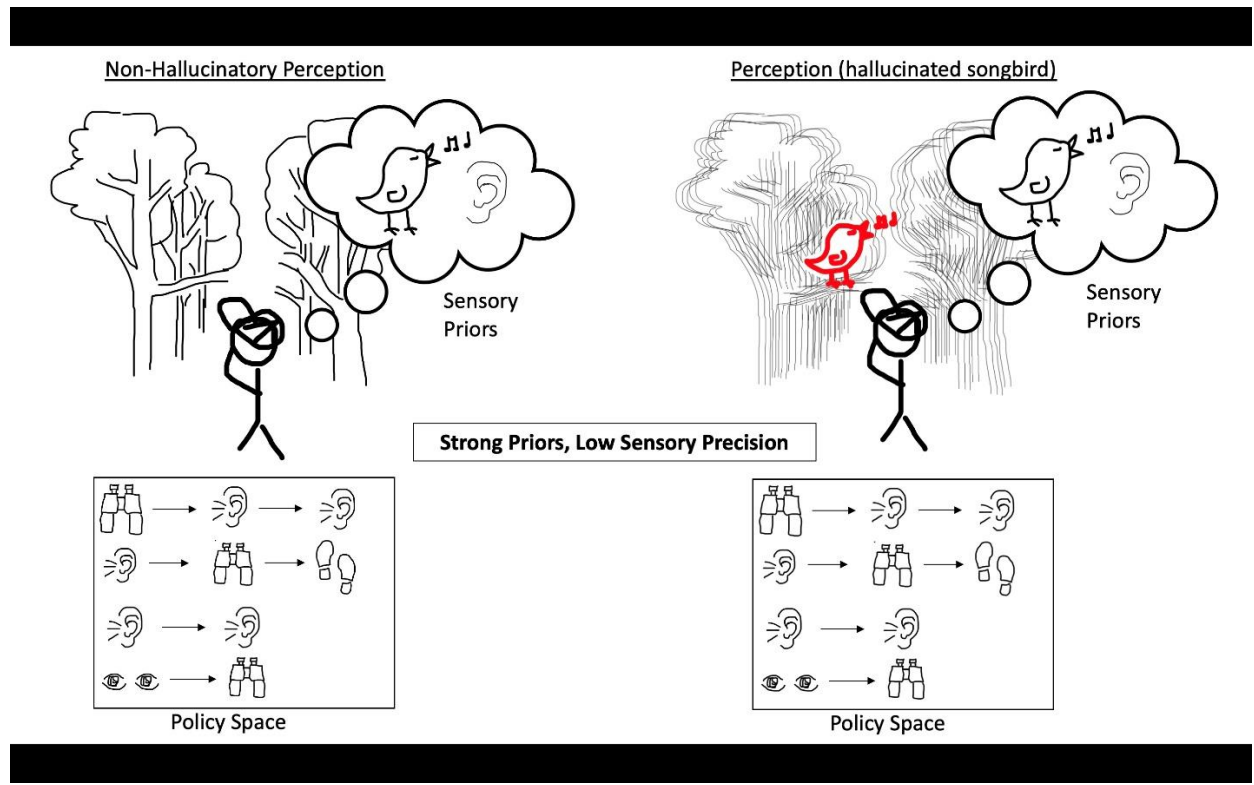

Figure 2:

On the left, the agent experiences reduced precision of incoming sensory evidence, but it has an expansive policy space without bias towards any string of actions and has equally weighted priors.

On the right, the agent experiences an increase in prior precision over policies, and their policy space is such that this results in having a bias towards actions that heavily involve listening and therefore the expectation of sound. This results in the hallucinated songbird because the agent is over-confident in these policies and less certain about other policies which may have implied other possible states of the world.

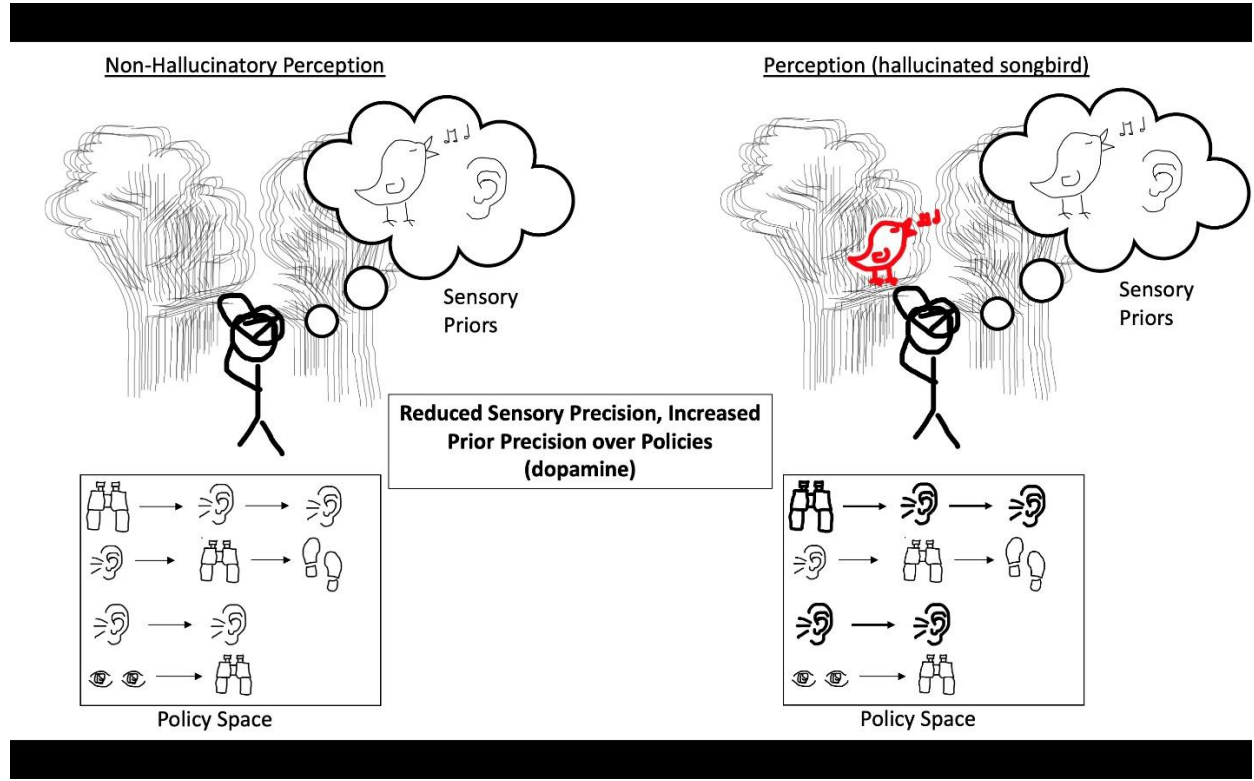

Figure 3:

On the left, the agent experiences incoming sensory evidence to which it ascribes low confidence, and it has equally weighted priors and policies in an expansive policy space. The result is an agent that does not hallucinate.

On the right, the agent experiences degraded precision of incoming sensory evidence and a decreased policy space. The decreased policy space now only includes a couple of policies that rely heavily on listening. The result is an agent that does not have a full range of policies adaptable to the world it finds itself in and which imply that a singing bird is present and thus the agent hallucinates the songbird.

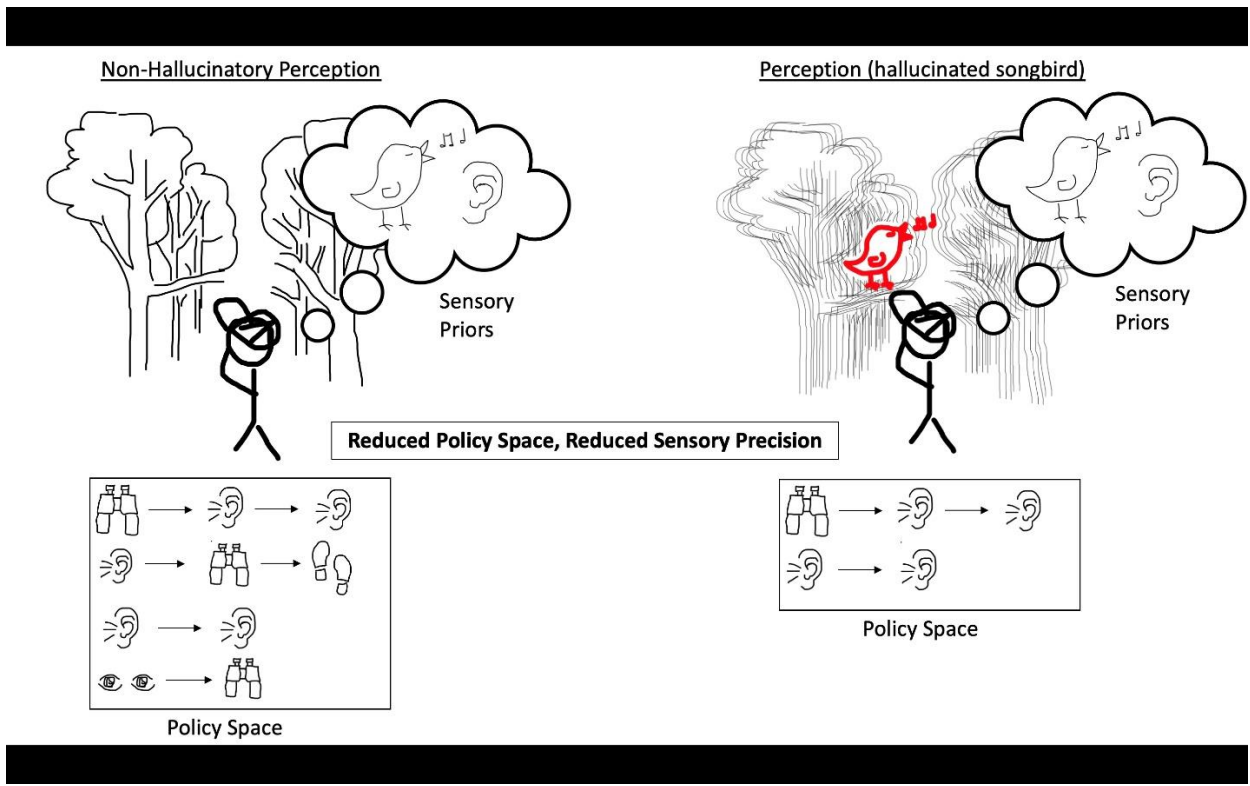

Figure 4:

On the left, the agent experiences degraded precision of incoming sensory evidence, a strong prior belief, a decreased policy space, and strong confidence in policy that relies exclusively on listening. This agent hallucinates the songbird as it would in the cases presented in figures 1-3.

On the right are four panels, each with a single change from the hallucinatory perception on the left. Each change represents an effect that treatment could have on a hallucinating agent (note that in almost all cases multiple parameters change with treatment, but for simplicity we demonstrate the possible changes independently).

The first is improved sensory precision: with clearer incoming sensory evidence, the agent can use its perception of the environment to correct for an over-weighted prior/policy or a maladapted policy space. We hypothesize that the approach of CSCs and ACT teams may help to increase the quality of and confidence in the external environment and as such improve the sensory precision.

The second is expanded policy space: with additional actions that the agent could take, it can make further observations about its environment and potentially choose a policy that less heavily implies the presence of the songbird, allowing it to correct for an over-weighted prior/policy and degraded sensory information. We hypothesize that this could occur as a result of cognitive training or therapy and social interactions and external support provided through ACT teams, or that policy space could be preserved via the early intervention provided by CSCs.

The third is reduced prior precision over policies: the agent now has equal confidence in the listening action as well as the listening, looking closely, and walking closer action, and as such a weaker prior favoring the presence of the bird. This is the effect hypothesized to be produced by dopamine blockade and helps the agent disengage.

The fourth is reduced influence of sensory priors: when the prior beliefs are weighted equally, the agent is not biased towards perceiving the songbird despite the limited policy space and strong prior precision over policies. We hypothesize that this could occur as a result of cognitive training or therapy and social interactions and external support provided through ACT teams, or that the formation of fixed maladaptive priors could be prevented via the early intervention provided by CSCs.

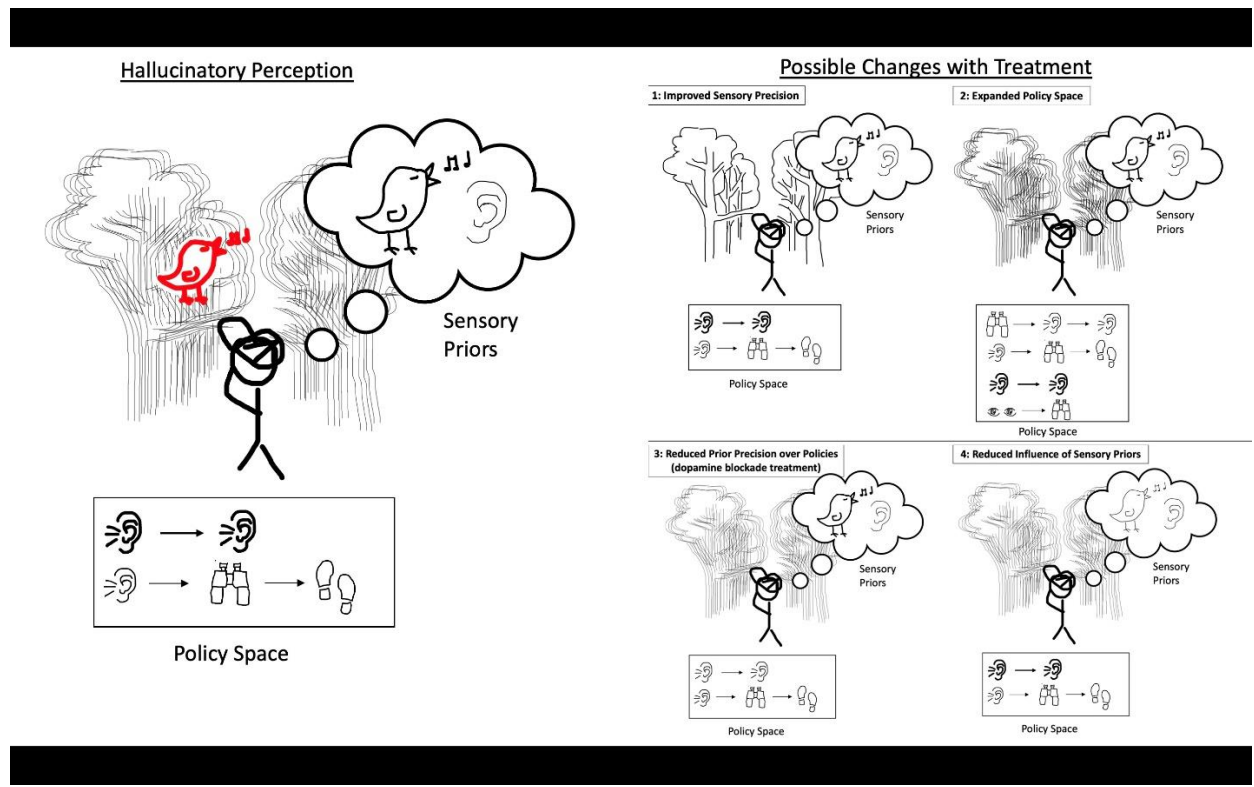

REFERENCES:
1. Bowie, C. R., & Harvey, P. D. (2006). Cognitive deficits and functional outcome in schizophrenia. Neuropsychiatric Disease and Treatment, 2(4), 531–536.